# Revealing Sharp Spectral Features with Complex-Frequency Excitations: Challenges and Opportunities

Jacob B Khurgin[1*], Andrea Alù[2], Vitaly Kozlov[2] and Philippe Lalanne[3]

[1]Deprtment of ECE, Johns Hopkins University, Baltimore MD 21218 USA

[2] Photonics Initiative, Advanced Science Research Center, City University of New York, New York, New York 10031, USA

[3]Laboratoire Photonique, Numérique et Nanosciences (LP2N), IOGS- Université de Bordeaux-CNRS, 33400 Talence cedex, France

[*]Corresponding author: jakek@jhu.edu

**Abstract**

Broadening of spectral and spatial responses due to intrinsic loss in real materials often hides sharp features. One recently recognized route to recover those features is to probe the system with complex-frequency (CF) signals that decay exponentially in time: a suitably tailored temporal decay can compensate for loss and reveal an intrinsic, narrow response. However, generating rapidly decaying optical waveforms in real time is often challenging (the required decay times may be in the range of tens of femtoseconds). A recently proposed alternative synthesizes the CF response numerically after detection of conventional, real-frequency signals using Fourier post-processing. Here we explore advantages and challenges of these approaches: we show that a physical CF excitation robustly sharpens spectral features in the presence of noise, while a post-detection synthesized CF response shows only limited improvement once realistic detection and readout noise is considered. At the same time, in low-noise conditions a much simpler post-detection filtering procedure attains equal or better recovery than the synthesized CF reconstruction, making the synthesis unnecessary in practice.

**Introduction**

As photonic technologies expand and enter many aspects of everyday life, their use in numerous applications is often hindered by optical loss. Unlike low-frequency signals (DC to microwave) that can be regenerated and amplified relatively easily, photons are far harder to restore because there is no simple photonic analogue of electrical bias. While some optical losses stem from scattering, imperfect coupling, and radiation, absorptive loss is the most pernicious: on a fundamental level it is linked to the medium's dephasing rate. Absorption not only removes energy (phonons) but also broadens sharp features, degrading resolution — not only in the spectral domain but also spatially, with especially serious consequences for sub-diffraction imaging and other high-resolution optical techniques.

It is no surprise that much effort has gone into mitigating the adverse effects of absorptive loss, and in recent years a noteworthy new approach has been proposed based on illumination with complex-frequency (CF) signals[1-4] . A CF signal at frequency $\tilde{\omega} = \omega - i\Gamma$ is at heart an exponentially decaying harmonic wave, $E(t) = E_0 e^{-i\omega t - \Gamma t} + c.c.$. When this signal is applied to any medium with a typical Lorentzian response $H(\omega) = A / (\omega_0^2 - \omega^2 - i\omega\gamma)$ the response becomes $H(\tilde{\omega}) = A / [(\omega_0^2 - \omega^2 + \gamma^2 - i\omega(\gamma - 2\Gamma)]$ with the material dephasing rate $\Gamma$ being partially or completely compensated or even reversed. Because a material's intrinsic decay appears as a commensurate decay of signals in devices made from that material, and because that decay is responsible for spectral and spatial broadening, an apparent reduction of the effective decay should improve the performance of many optical devices. The effect is potentially most valuable for systems built from highly lossy, strongly dephasing materials — metals, with their sub-picosecond dephasing times, being the most prominent example[5]. For instance, because imaging resolution with metallic superlenses is severely limited by dephasing [6-9], a reduction of dephasing could move superlensing from an exotic concept toward a practical technique

The concept of CF loss compensation has been verified experimentally using acoustic waves with frequencies in the hundreds of hertz[4], and it appears that the required waveforms can be reliably synthesized in the microwave range. However, at optical frequencies[10], generating waveforms that decay exponentially faster than a few hundred femtoseconds—which is the typical range of damping constants in metals and semiconductors—poses a significant challenge. Achieving this feat with a reasonable fidelity is not a simple task. Furthermore, since waveforms corresponding to complex frequencies cannot extend indefinitely in time, the initial conditions inevitably generate a transient response that can interfere with the useful signal oscillating at the complex frequency, necessitating the use of careful dispersion engineering, time gating or post-processing techniques to isolate the desired information, when possible.

The idea of virtually compensating losses (or, more accurately, dephasing) in the optical range would remain largely theoretical if not for recent innovative proposals that avoid introducing complex frequencies within the physical or real domain[11-14]. The key concept is simple: synthesize the system's response to signals at complex frequencies by combining its measured responses at many real-frequency inputs via Fourier synthesis. In the rest of the apaper we shall refer to this approach as a synthesized CF (SCF) technique to distinguish it from the conventional or generated in real time CF (RTCF) approach when the CF signal is generated in physical domain prior to detection. The mathematical apparatus of SCF is straightforward. Let $H(\omega)$ denote the system's frequency response (which may be absorption, scattering, or transmission). Assuming linearity and performing inverse Fourier transform of $H(\omega_1)/(\omega - \omega_1 - i\Gamma)$ is expected to yield what would be a temporal response to a signal with complex frequency $\tilde{\omega}_1 = \omega_1 - i\Gamma$, that decays with rate $\Gamma$ in addition to the resonant response of the system at a set of eigenfrequencies $\omega_{0,n}$, decaying at their own rates $\gamma_n$. Assuming that the natural decay times $\gamma_n$ are faster than $\Gamma$, by applying time gating in the computed time domain we can then isolate the component corresponding to the desired complex-frequency excitation. This procedure can effectively reduce the apparent spectral broadening in $H(\omega)$. Yet, it incurs in a steep energy penalty because only the late, exponentially small tail of the decaying signal can be used.

Figure 1 illustrates the two approaches: in the physical real-time domain (Fig. 1a), an RTCF excitation is created up front — for example by a pulse shaper acting on real-frequency harmonics — and the tailored waveform interacts with the sample whose spectral response is *H(ω).* The detector then records the time-gated portion of the transmitted/reflected signal, where CF-induced spectral narrowing is most pronounced.

By contrast, in the SCF scheme (Fig. 1b), the sample is illuminated with conventional real-frequency harmonics (simultaneously or sequentially) and the detector measures the usual response $H(\omega)$ or $H(\lambda)$ (typically as integrated charge per frequency channel). That measured spectrum is then processed digitally after detection to synthesize the equivalent CF time trace and recover a CF-like response. Crucially, in SCF, the recorded signal — and noise that accompanies it — is processed post-detection; as we shall show in the following, that this issue can lead to undesirable effects.

The SCF approach has been shown to improve superlens resolution in the microwave and IR regions[14], resolve molecular resonances [13] and most recently to "reveal" plasmonic resonances normally obscured by broadening[12]. It therefore appears that post-detection synthesis of CF responses could be applied wherever broadening limits performance. This picture, however, invites skepticism. Many powerful post-detection processing methods already exist in the time, space, and frequency domains and can substantially enhance apparent resolution[15], yet none can overcome the fundamental limits imposed by noise. It appears overly optimistic to expect that the SCF technique can evade these constraints or outperform well-established post-processing techniques in realistic noisy settings. In this article we examine exactly these questions by comparing three approaches: (1) the physical (real-time) RTCF excitation, (2) the SCF reconstruction, and (3) a standard inverse-filtering post-processing[16, 17]. We apply these approaches to the basic task of recovering sharp spectral features obscured by broadening; although elementary, this test case captures the essential trade-offs, and the conclusions carry over to related problems such as improving the resolution of superlenses.

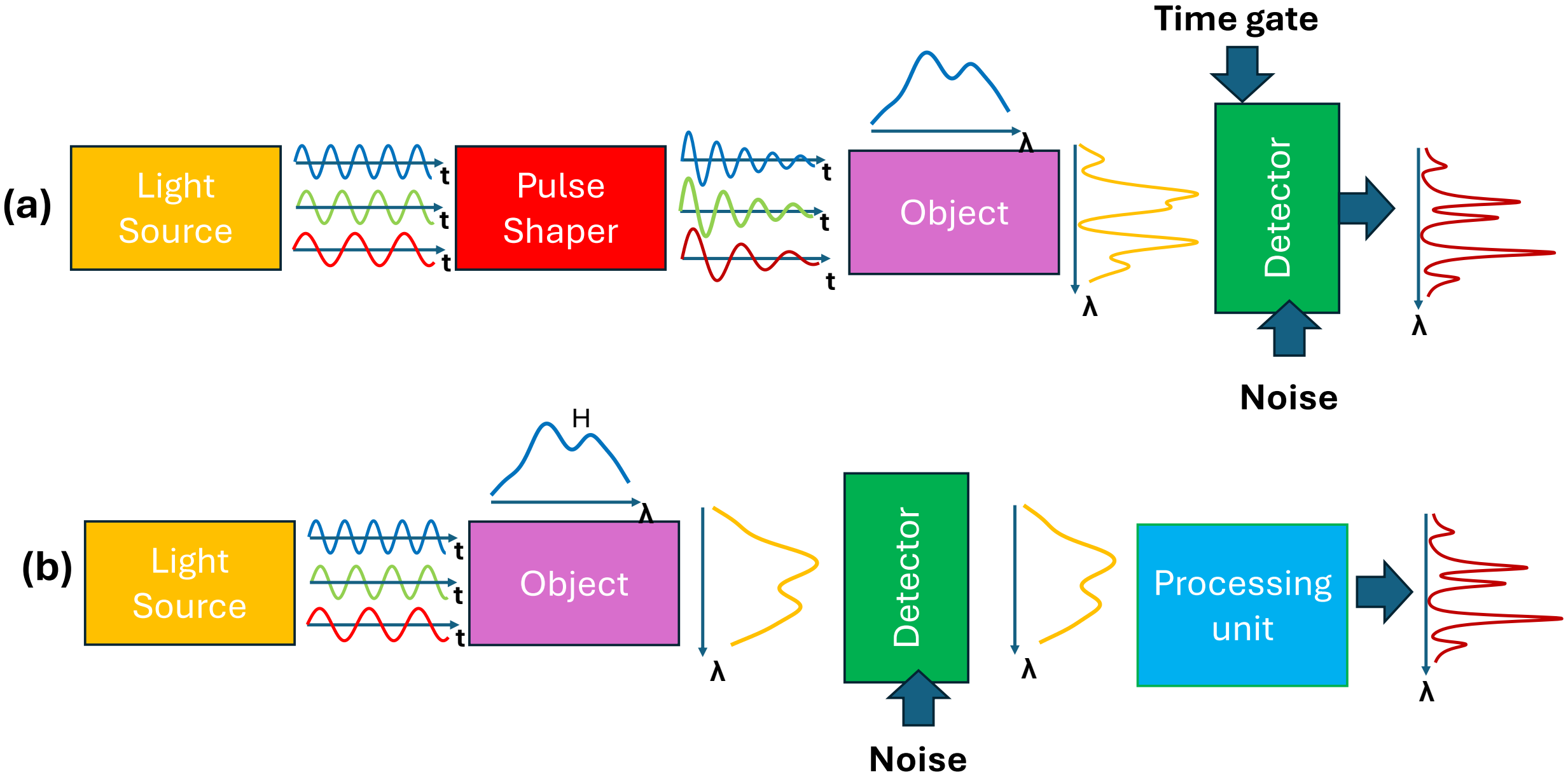

**Figure 1.** Schematics of two CF approaches to improving spectral resolution. (a) Real-time-CF (RTCF) technique — the CF excitation is produced before interaction with the object, and the detector records the object's direct response to that CF input. (b) Synthesized-CF (SCF) technique — the sample is illuminated with real-frequency inputs and the CF response is synthesized digitally after detection; this is, in practice, just one of many post-detection filtering or reconstruction methods and offers no fundamental distinction from other computational filtering techniques.

**Performance in noiseless environment**

We start with the most basic model of a system with a Lorentzian resonance at $\omega_0$ $H(\omega) = A/(\omega_0^2 - \omega^2 - i\omega\gamma)$ where $A$ can be a scattering, reflection or transmission[12, 13] Its power spectrum $|H(\omega)|^2$ shown in Fig. 2a, dashed line, with frequency scaled to $\omega_0$. In the examples that follow we focus on a plasmonic mode resonance, for which loss is the dominant issue to be mitigated. For concreteness we choose $\omega_0 = 2\pi \times 400THz$ (corresponding to wavelength of 750nm) and $\gamma = 1/10 fs$ which is a typical value for noble metals giving us $Q = \omega_0/\gamma \approx 25$ We then apply a CF signal at $\tilde{\omega} = \omega - i\Gamma$ $E(t) = E_0 e^{-i\omega t - \Gamma t} + c.c.$ whose FT is $E_0/(\omega - \omega_1 - i\omega_1\Gamma)$. The response to this signal is

$$S(t) = -\frac{AE_o}{2\pi}\int_{-\infty}^{\infty}\frac{e^{-j\omega_1 t}}{(\omega-\omega_1-j\Gamma)(\omega_0-\omega_1-j\gamma/2)(\omega_0+\omega_1+j\gamma/2)}d\omega_1 = S_{\tilde{\omega}}(t) + S_0(t) \tag{1}$$

where the first term

$$S_{\tilde{\omega}}(t) = \frac{AE_0}{j}\frac{e^{-j\omega t-\Gamma t}}{\left[\omega_0-\omega \ -j(\gamma/2-\Gamma)\right]\left[\omega+\omega_0+j(\gamma/2-\Gamma)\right]} \approx \frac{AE_0}{2j\omega_0}\frac{e^{-j\omega t-\Gamma t}}{\omega_0-\omega \ -j(\gamma/2-\Gamma)} \tag{2}$$

is the desired CF response (usually referred as “forced”) while the second component

$$S_0(t) = -\frac{AE_0}{2j\omega_0}\left[\frac{e^{-j\omega_0 t-\gamma t/2}}{\omega_0-\omega-j(\gamma/2-\Gamma)} + \frac{e^{j\omega_0 t-\gamma t/2}}{\omega_0+\omega+j(\gamma/2-\Gamma)}\right] \tag{3}$$

represents the transient (or “free”) oscillations at resonant frequency $\omega_0$ that obscures the improvement in resolution. In Fig.2a,b we show a typical response and its components for $\omega = \omega_0 + \gamma/2$ and $\Gamma = 0.6\gamma/2$.

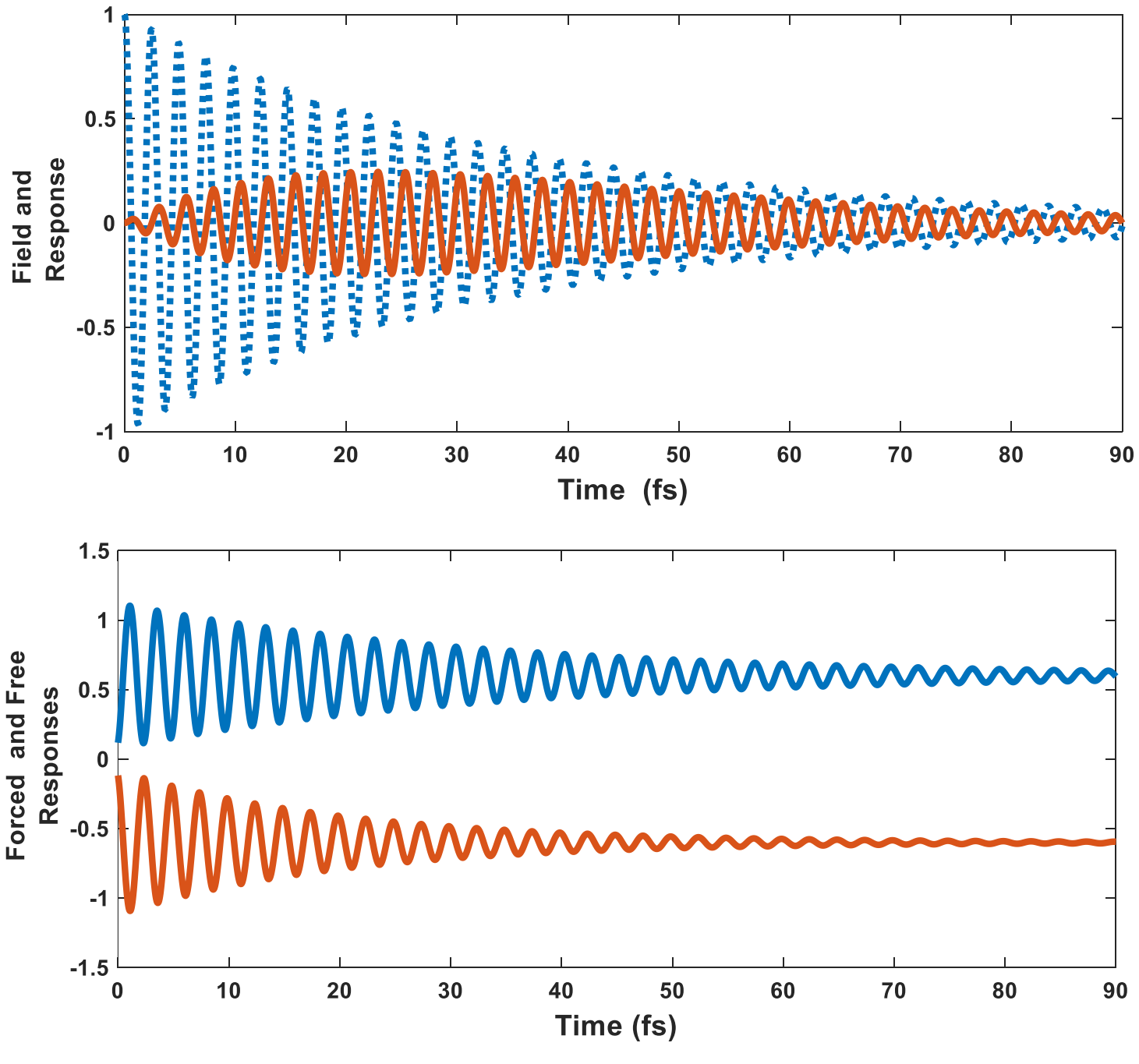

**Figure 2.** A CF signal (dashed line) and the system response to it (solid line).. (b) Forced (upper curve) and free (lower curve) components of the response

If we divide the total energy of the forced response (2) by the energy of the CF input signal, we shall obtain

$$\frac{\int_0^\infty |S_{\tilde{\omega}}(t)|^2\,dt}{\int_0^\infty |E(t)|^2\,dt} = \frac{A^2}{(\omega_0^2-\omega^2)^2+\omega^2(\gamma-2\Gamma)^2} = |H(\tilde{\omega})|^2 \tag{4}$$

i.e., we precisely retrieve the desired response of the transfer function at theCF $\tilde{\omega}=\omega-i\Gamma$ signal. Actually almost the same result can be obtained by looking at the energy of free response (3). Unfortunately one cannot untangle two responses which hinders the result – We can address this problem by performing the integration only along the tail of the pulse, starting from some onset time $\tau$, after which the contribution from the natural resonances of the system can be reduced:

$$|H(\tilde{\omega},\tau)|^2 = \frac{\int_\tau^{t_{max}} |S(t)|^2\,dt}{\int_\tau^{t_{\max}} |E(t)|^2\,dt} \tag{5}$$

Naturally, the best result is obtained at $\tau >> \gamma^{-1}$ when the free response has decayed but the forced one still remains [1]. However, as $\Gamma$ approaches $\gamma/2$ the forced response also drops dramatically, so the achievable resolution becomes increasingly limited by noise. The situation can be somewhat improved by using higher order CF signal $E_n(t)=E_0t^ne^{-i\omega t-\Gamma t}+c.c.$ but the improvement is marginal[1, 18].

To illustrate this challenge, we first consider the complex frequency response in the absence of noise. As an example, we use $\Gamma=0.9\gamma/2$: Figure 3a shows the response to RTCF for five different onset times from 0 to 200fs and $t_{\max}=250fs$, using the analytic expressions (2) and (3). Figure 3b shows the same calculation performed with the SCF approach, i.e., post-processing the calculated spectra at real frequencies to reproduce the CF excitation synthesized using (1). As expected, aside from small ripples originated from the finite frequency step in (1), the two methods agree well, and it is apparent that the effective bandwidth is gradually reduced as the onset time $\tau$ increases. Even without time-gating ($\tau$=0) the width of the resonance decreases. The reduction of the effective bandwidth is limited by the temporal extent of integration $t_{\max}$. In these calculations, the post-processed approach appears to provide an advantage: it enables improved resolution without the need to modulate the temporal waveform and generate short, high-fidelity exponentially shaped pulses. To understand why, consider the time–frequency Fourier relationship. As the spectral response in Fig. 3b narrows, its Fourier-conjugate temporal response must broaden. Truncating the time integration before a given onset effectively removes the earliest portion of the temporal response. In the time domain this is equivalent to applying a high-pass filter with a cutoff at the chosen onset time.

Although this may seem counterintuitive—one often thinks of filtering in the frequency domain—our goal is to reduce frequency-domain broadening. Because time and frequency are reciprocal of each other, filtering the temporal response is the correct way to shape the spectral response. By selectively excluding the early (mostly free) temporal contributions, the synthetic CF method narrows the effective spectral width and thus improves resolution.

But is this the easiest or the best way to achieve high pass filtering to recover sharp resonances? Obviously not as the best filter is the inverse matched filter, i.e. the temporal response $1/|h(t)|$ which means multiplying the temporal response $s(t)$ (obtained by applying inverse Fourier transform to the spectral response) by rising exponential $e^{+\Gamma t}$ for reduction of broadening. It is easy to see that since temporal(impulse) response in our example $h(t)=e^{-i\omega_0 t-\frac{1}{2}\gamma t}$, the new response will be $h'(t)=e^{-i\omega_0 t-\left(\frac{1}{2}\gamma-\Gamma\right)t}$. In fact, we can write for the general response $H(\omega)$ the expression for the spectral response obtained with multiplying its Fourier conjugated temporal response by $e^{+\Gamma t}$

$$H'(\omega)=\int_0^{t_{\max}\to\infty} h(t)e^{\Gamma t}e^{-i\omega t}dt \to H(\omega-i\Gamma) \tag{6}$$

(as long as $h(t)e^{i\Gamma t}$ remains integrable). Thus, one obtains exactly the desired response at the complex frequency $H(\tilde{\omega})$, free from spurious transient contributions and computed with far less effort — simply by applying the Fourier transform pair to the measured spectrum with an intermediate multiplication by the compensating exponential factor.

This behavior is illustrated in Fig. 3c for the same parameter set used earlier, across five integration times $t_{\max}$ from 50 fs to 250 fs. As expected, the spectral response becomes progressively narrower with increasing integration time, and the amount of narrowing exceeds that seen for the CF excitations in Figs. 3a and 3b. Matched filtering is also far less computationally demanding than the SCF approach, calling into question the practical utility of the latter.

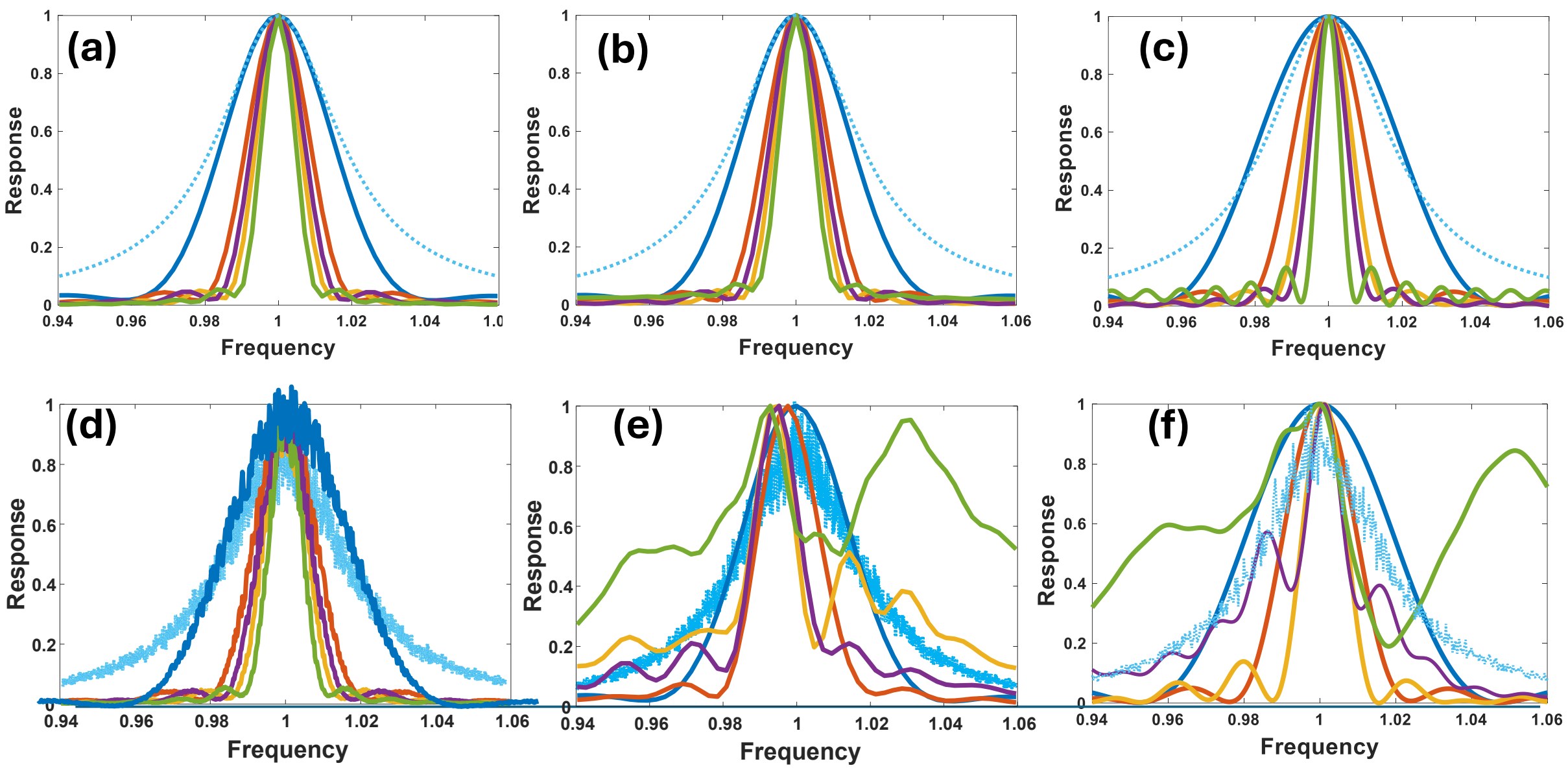

**Figure 3**. Restoring a sharp resonance using different CF excitation approaches. (a) Response to a real-time CF (RTCF) excitation. The frequency axis is normalized to the resonance at 400 THz. The blue dashed line shows the response to a real-frequency excitation; solid lines show responses to CF excitations constructed by time-domain integration starting at onset times of 0, 50 fs, 100 fs, 150 fs and 200 fs and extending to a maximum time of 250 fs. (b) CF response synthesized after detection (post-detection reconstruction). (c) Inverse (matched) filtering applied to the real-frequency response, using different sampling windows in the time domain with onsets at 0, 50 fs, 100 fs, 150 fs, 200 fs and 250 fs. (d)–(f) Same as panels (a)–(c), respectively, but with detector noise included. In these noisy panels the response to the real-frequency excitation is shown as the jagged blue. In all panels the frequency is normalized to the resonant frequency and the magnitude of response is normalized to 1 (indicating the same detected power for all cases)

## Impact of noise

As demonstrated above, both the SCF and inverse-filtering approaches work by emphasizing the late (trailing) portion of the system temporal response relative to its immediate response. Because that temporal response decays rapidly, at some point it is expected to fall below the noise level. We can define an SNR-threshold time $t_{SNR}$, or the effective observation time, after which we cannot expect to distinguish the signal from the noise. $t_{SNR}$ ultimately limits the attainable spectral resolution: the minimum resolvable feature scale is set by the inverse of the time extent over which the Fourier-conjugate signal remains above the noise floor. This conclusion is general for any post-detection method that seeks to enhance resolution in space, time or frequency.

Let us now consider what are the possible sources of noise. We assume the spectrum under real-frequency excitation is obtained in one of two common ways: either a broadband excitation is dispersed (typically with a grating) onto a detector array, or a narrowband tunable source is stepped through frequency while a single detector records the signal sequentially. In either case, various noise sources affect the measured spectrum.

Temporal fluctuations such as shot noise and thermal noise are present in the photocurrent, but because most spectroscopic readouts integrate this current into a total exposure (usually measured as accumulated charge), instantaneous time-domain fluctuations are largely averaged out. What remains is pixel-to-pixel noise in the integrated signal at each real frequency channel. A primary contribution is pixel-to-pixel shot noise: the total exposure at a given real frequency is proportional to the number of detected photons at each pixel, so photon-counting statistics introduce an irreducible variance that appears as a ripple on the recorded spectrum (Fig. 3d–f)[19]. When a detector array is used, pixel-to-pixel variations in responsivity, gain, and readout behavior also produce fixed-pattern and calibration-type errors. These spatial nonuniformities are conceptually similar to relative intensity noise (RIN) in the time domain: both act multiplicatively, modulating the measured amplitude and producing ripples across the spectrum. A tunable source can introduce comparable amplitude and frequency instabilities, which likewise manifest as spectral ripple.

A common feature of these contributions is that their magnitude grows with the detected signal (total accumulated charge). There are also signal-independent, additive noise sources — for example dark current and stray light . In addition to this there is also noise due to error in retrieving the phase spectrum[20]. which is necessary to post process the signals in the SCF approach. The combined effect of all contributions is the noisy, rippled spectrum illustrated in Fig. 3d–f. for the case when signal-dependent noise dominates. It is the balance between signal-dependent and signal-independent noise that ultimately determines whether the real-time CF and SCF approaches are affected similarly by noise or whether one approach offers a practical advantage.

.

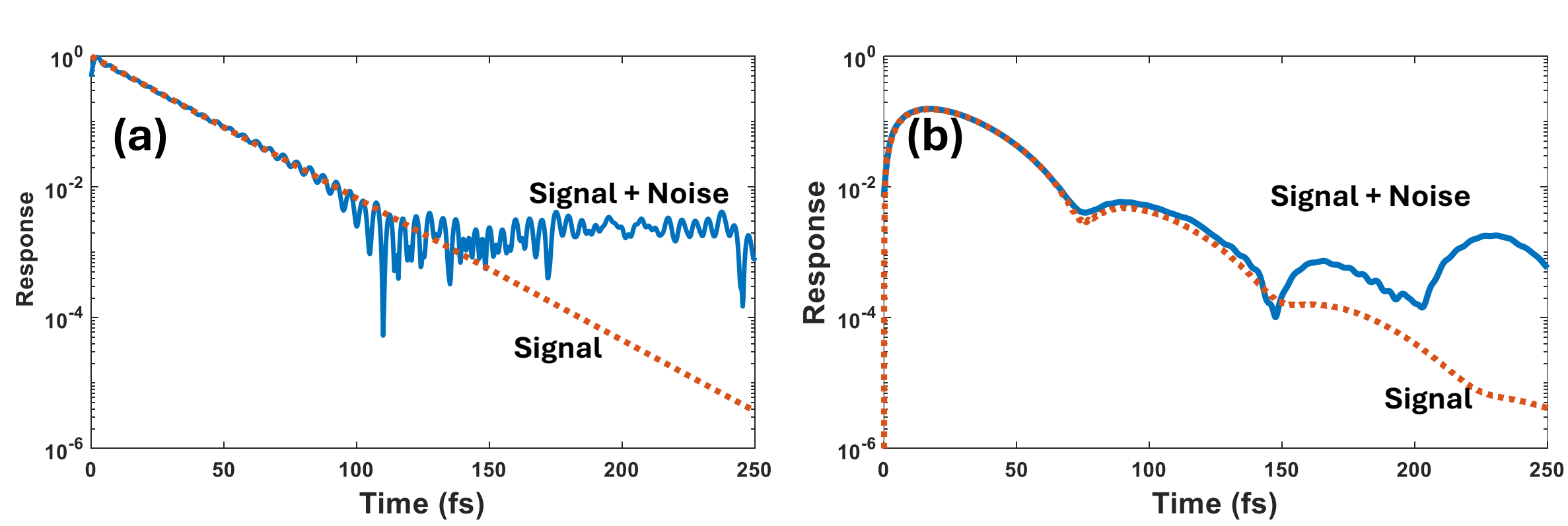

**Figure 4.** Computed time-domain responses and SNR limits for post-detection methods. (a) Power of the temporal response from the inverse FT of the noisy spectrum (solid) with noiseless signal (dotted); noise exceeds signal at $t_{SNR}$ ≈ 120 fs. (b) Computed temporal response for the synthesized CF case with added noise; noise dominates at times exceeding 140 fs.

Figure 4a (solid line) shows the temporal response $|h'(t)|^2$ computed through the inverse Fourier transform of the noisy spectrum in Fig. 3f. The noise is effectively white in the time domain, and it exceeds the signal (dotted line) at $t_{SNR} \approx 120$ fs. Consequently, when this temporal waveform is multiplied by the rising exponential in Eq. (6), the amplified noise at large times will dominate. As a result, performing the Fourier transform in Eq. (6) with $t_{max} > t_{SNR}$ will not recover the expected narrowed spectrum. Indeed, Fig. 3f shows this effect clearly: integration up to about 150 fs produces a narrowed feature, but longer integration times yield broad, featureless shapes that no longer resemble a resonance, clearly seen in noiseless case of Fig.3c Thus, inverse filtering is fundamentally limited by noise.

What about the SCF approach? Figure 4b shows $|S(t)|^2$-the computed response to the CF signal with added noise, $|S(t)|^2$ and indicates that noise dominates the signal at $t_{SNR} \approx 140$ fs. Consequently, when only the tail of the response is used to compute the spectrum via (5), the result (Fig. 3e) deviates substantially from the noiseless case (Fig. 3b). The achieved spectral narrowing is no better than that obtained with the inverse-filtering method (Fig. 3f). Given that the SCF method is also more computationally intensive than inverse filtering, its use is difficult to justify.

Next, we analyze how the presence of noise affects the performance of real-time CF excitations. Since CF signals are generated prior to detection, as long as the noise is signal-dependent, the shot and RIN noises will be added at detection to the already narrowed spectrum generated by time gating of the temporal response $|S(t)|^2$. Hence, the narrowing achieved in the noiseless case will be largely preserved in the real-time excitation scenario, as shown in Fig.3d. We can further clarify the difference between the real-time and post-processed CF approaches: in both methods the spectral response is ultimately inferred from the tail of the temporal response, but the noise that contaminates that tail behaves differently.

With a real-time CF illumination, the measurement is effectively time-gated to the late-time portion of the response: only the signal (and its associated noise) within that selected time window contributes, while noise from earlier times is largely excluded. Consequently, shot noise and RIN affecting the gated window scale with the detected energy actually present there. If, however, the dominant noise sources at long times are dark current, or stray light, the SNR will still degrade and the real-time approach will offer no advantage over SCF.

By contrast, in the SCF method the signal is first recorded over the full frequency band and then transformed into the time domain. Detection noise therefore reflects the total detected energy; when mapped into the computed time trace this noise is approximately white, so its amplitude at long (computed) times can exceed the decaying signal. That elevated noise level erodes any spectral-narrowing benefit obtained by emphasizing the response tail.

**Revealing the resonances using RTCF, SCF, and inverse filtering**

We now apply the inverse-filtering approach to the task of “cleaning up” a spectrum that contains multiple overlapping broad resonances, similarly to what was attempted with the SCF method[12, 13]. As an example, consider three resonances of different strengths and widths that produce the power spectrum shown as dashed curve in Fig. 5a-c, where the individual peaks are not resolved.

First, we consider RTCF illumination. As shown in Fig. 5a, the three resonances of differing strength are readily resolved. As discussed above, as long as shot noise and RIN-like contributions dominate, the resolution for real-time excitations remains robust because the time-gated measurement contains only the weak, late-time signal and its correspondingly small detection noise.

Applying the SCF procedure to the same data produces comparable enhancement in the noise-free case (Fig. 5b, curve labeled "No Noise"). However, after adding shot noise and RIN the enhancement disappears: the "Noisy" curve shows the resonances are completely lost.

Figure 5c displays results for inverse-filtering methods. Like SCF, inverse filtering recovers the three resonances in the absence of noise but is similarly defeated by shot noise and RIN in the noisy case. If anything, inverse filtering gives slightly better peak visibility than the SCF reconstruction under the same conditions.

More generally, though, the conclusion is straightforward: if a clean, noise-free spectrum composed of overlapping Lorentzian lines is available (and especially if phase information is also known), decomposing it into its constituent resonances is a routine task. Numerous well-established fitting and deconvolution methods [21-23] can recover the individual Lorentzian (and not only Lorentzian) spectra reliably, and the synthesized-CF approach is by no means the most efficient or effective option.

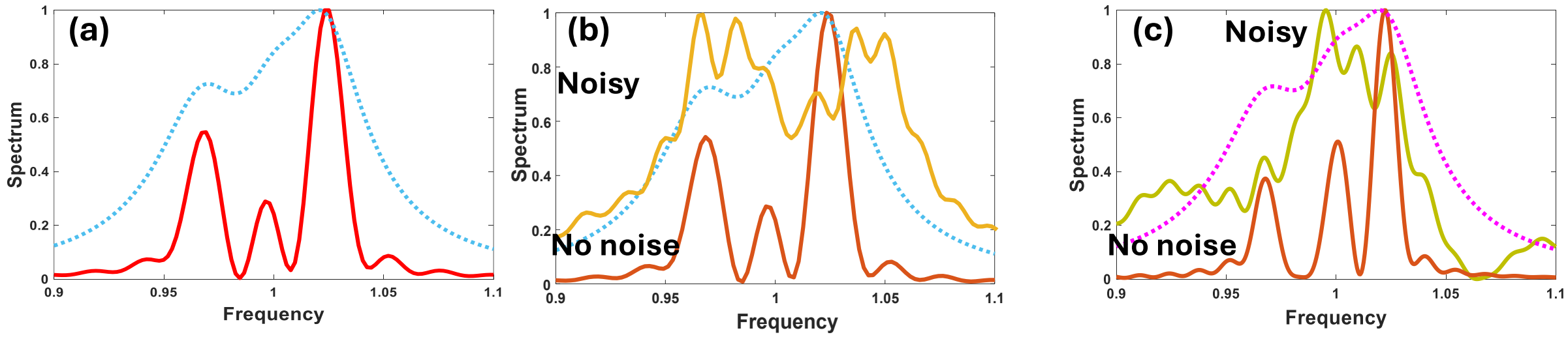

**Figure 5.** Resolving overlapping resonances. Dashed: original unresolved power spectrum. Panels compare methods and show noise-free and noisy cases, noise being signal dependent (a) RTCF illumination: resonances recovered in both cases. (b) SCF reconstruction: recovered only without noise; (c) Inverse filtering: similar to SCF — recovery in the noise-free case, obscured by noise (slightly better peak visibility than SCF under the same noisy conditions).

Before moving on, it is worth scrutinizing the common phrase "loss compensation" used to describe the effect of CF illumination on resolution. Narrowing a Lorentzian line shape under CF illumination does not reduce the total absorption — the integrated area under the absorption curve remains unchanged (the spectra in Figs. 3 and 5 are normalized and therefore do not display that conservation). Moreover, CF illumination can also sharpen features when the line broadening is inhomogeneous (e.g., Gaussian), a mechanism that is unrelated to absorption strength. And last, but not least, reducing absorption by say optical pumping and even inversing it to achieve gain does not affect the linewidth. For these reasons, we suggest the term "broadening compensation" is more appropriate than "loss compensation," since the effect acts to counteract spectral broadening rather than to alter total loss.

**Complex frequency and spatial resolution**

The CF method has been shown to enhance resolution not only in the spectral domain but also in the spatial domain—for example, by mitigating the loss (or, more precisely, the broadening) of surface plasmon (or phonon) polaritons in a metal or dielectric superlens. It is plausible that the same idea may improve resolution in the temporal domain. In any case, regardless of the domain in which one seeks to enhance sharp features after detection, the procedure amounts to applying some form of high-pass filtering in the Fourier-conjugate domain. Many digital filtering techniques exist, each with its own advantages and drawbacks. In this work we have shown that post-detection synthesized CF is just one such technique; based on our results, it is not necessarily superior to simpler approaches. In particular, synthesized CF offers no improvement over the basic inverse matched-filter method while being substantially more computationally demanding.

While a full analysis is beyond the present scope, similar considerations apply to spatial resolution enhancement with CF excitations — for example, in a superlens whose performance is degraded by the losses associated with the negative permittivity required for superlensing [8, 24]. Under real-frequency illumination the point-spread function (PSF), $h(\boldsymbol{r},\omega)$ (with $r$ the in-plane coordinate), is washed out; equivalently, the spatial-frequency response or modulation transfer function (MTF), $H(k,\omega)$ (with $k$ the in-plane wavevector), falls off beyond some cutoff $k_{\max}(\omega)$, which limits the achievable resolution to roughly $1/k_{\max}$. Under illumination with complex frequency $\tilde{\omega}$ the MTF $H(k,\tilde{\omega})$ broadens to $k_{\max}(\tilde{\omega}) > k_{\max}(\omega)$ and thus increasing the resolution [4, 18]. If one illuminates the superlens with a wave $s(k,\omega) = E_k e^{i(kr-\omega t)} + c.c.$ then the image pattern will be $H(k,\omega)s(k,\omega)$ Performing the same linear transformation as in (1) yields

$$s'(k,t) = -\frac{E_k}{2\pi}\int_{-\infty}^{\infty}\frac{H(k,\omega_1)e^{-j\omega_1 t}}{(\omega-\omega_1-j\Gamma)}d\omega_1 = S_{\tilde{\omega}}(k,t) + S_0(k,t) \tag{7}$$

where the first term is the forced response to the CF excitation with frequency $\tilde{\omega} = \omega - i\Gamma$ and the second term describes the turn on oscillations at the slab plasmon (or phonon)–polariton mode frequencies — i.e., the poles of $H(k,\tilde{\omega})$. Appropriate time gating and subsequent integration suppress the transient pole contributions and isolate the desired CF response, $E_k H(k,\tilde{\omega})$. Compared with the real-frequency image the CF response enhances high spatial frequencies relative to low ones and therefore improves resolution.

All said and done, synthesizing the complex-frequency response by linearly combining real-frequency images and then gating in the computed time domain effectively amounts to passing the real-frequency image through a high-pass spatial filter, extending the usable spatial bandwidth from $k_{\max}(\omega)$ to $k_{\max}(\tilde{\omega}) > k_{\max}(\omega)$. Spatial filtering and deblurring are mature well-trodden fields with many established computational techniques[25, 26] — some not even requiring phase information, unlike the synthetic-CF approach — yet all are ultimately noise-limited, and the synthesized-CF method is no exception. The image detected at real frequencies is sampled on discrete pixels, and detection introduces approximately white noise (shot noise plus RIN-like contributions and pixel nonuniformities) that contaminates all spatial frequencies $k$. Similarly, noise is introduced when the near-field image is scanned by a tip with sub-wavelength aperture. Whenever this noise exceeds the signal at high $k$, no amount of mathematical processing can recover those components from beneath the noise floor, so the resolution cannot be improved.

And, just as shown above for spectral resolution, when the CF field is generated physically before detection the time-gated image excludes most of low spatial frequency energy that would otherwise contribute to detection noise. Consequently, the time-gated RTCF image is less prone to producing white-noise contamination at high spatial frequencies, improving the prospects for genuine resolution gain compared with post-detection SCF procedures. Of course, if the dominant source of noise is dark current, or background, there is no difference between synthesizing CF signal in physical or computational domain, or for that matter using one of alternative computational deblurring methods.

## Conclusions

In this note we examined the use of CF illumination to resolve sharp resonances in the spectral response of lossy materials and devices. When the CF excitation is generated physically (pre-detection) and the signal is time-gated in real time after detection, resolution can indeed improve — limited, of course, by background and detector noise sources such as dark current and stray light. Importantly, with physical CF the shot noise and pixel-related noise originate only from the relatively weak, time-gated useful signal, so those photon- and detector-limited noise terms scale with the small detected energy in the gate interval and are therefore reduced compared with the full-trace case.

By contrast, synthesizing CF responses post-detection is essentially a form of post-detection filtering: it offers no fundamental advantage over established computational deblurring or filtering methods while adding substantial computational complexity. The synthesized-CF method is also vulnerable to shot noise, RIN-like fluctuations, and pixel nonuniformities that stem from the total detected energy; when these noise terms are mapped into the computed time they overwhelm the useful signal and cap any achievable resolution gain.

When comparing the two approaches to realizing a CF excitation, one important practical distinction should be noted: post-detection synthesis requires access to the full complex field (amplitude and phase) — either via coherent detection or a reliable phase-retrieval step — because the procedure reconstructs a time-domain waveform from measured spectra. By contrast, generating the CF excitation in the physical domain can be relatively straightforward and does not demand post-hoc phase reconstruction: for example, injecting a short pulse into a Fabry–Pérot cavity (analogous to ring-down spectroscopy[27]) produces the desired exponentially decaying waveform directly.

Similar conclusions apply to spatial-resolution enhancement, for example with superlenses. If the dominant noise sources are thermal noise, dark current, or background light, neither a physically generated CF excitation nor a synthesized-CF post-detection approach provides an advantage over established deblurring and filtering techniques — and the synthesized-CF method is significantly more computationally demanding. If the dominant noise is shot noise or pixel-related noise, a physically generated CF excitation can offer a relative improvement because the time-gated measurement contains only the weak, useful signal and its associated (reduced) detection noise. Whether that potential gain justifies the additional experimental complexity of implementing physical CF illumination remains an open question.

Overall, CF techniques are a relatively fresh and intriguing direction; they may not become universally practical in every setting, but we are cautiously optimistic they will find useful applications where their unique advantages outweigh the added complexity.